# Optimal cellular mobility for synchronization arising from the gradual recovery of intercellular interactions

Running head: Optimal cellular mobility for synchronization

**Koichiro Uriu** [1,2,3,7], **Saúl Ares** [3,4,6], **Andrew C. Oates** [2] **and Luis G. Morelli** [2,3,5,7]

1. *Theoretical Biology Laboratory, RIKEN Advanced Science Institute, 2-1 Hirosawa, Wako, Saitama, 351-0198, Japan*
2. *Max Planck Institute of Molecular Cell Biology and Genetics, Pfotenhauerstr. 108, 01307 Dresden, Germany*
3. *Max Planck Institute for the Physics of Complex Systems, Nöthnitzer Str. 38, 01187 Dresden, Germany*
4. *Grupo Interdisciplinar de Sistemas Complejos (GISC)*
5. *Departamento de Física, FCEyN, UBA, Ciudad Universitaria, 1428 Buenos Aires, Argentina*
6. *Present address: Logic of Genomic Systems Laboratory, Centro Nacional de Biotecnología - CSIC. Calle Darwin 3, 28049 Madrid, Spain*
7. Authors to whom any correspondence should be addressed.

E-mail: uriu@mpi-cbg.de and morelli@mpi-cbg.de

**Abstract**

Cell movement and intercellular signaling occur simultaneously during the development of tissues, but little is known about how movement affects signaling. Previous theoretical studies have shown that faster moving cells favor synchronization across a population of locally coupled genetic oscillators. An important assumption in these studies is that cells can immediately interact with their new neighbors after arriving at a new location. However, intercellular interactions in cellular systems may need some time to become fully established. How movement affects synchronization in this situation has not been examined. Here we develop a coupled phase oscillator model in which we consider cell movement and the gradual recovery of intercellular coupling experienced by a cell after movement, characterized by a moving rate and a coupling recovery rate respectively. We find (1) an optimal moving rate for synchronization, and (2) a critical moving rate above which achieving synchronization is not possible. These results indicate that the extent to which movement enhances synchrony is limited by a gradual recovery of coupling. These findings suggest that the ratio of time scales of movement and signaling recovery is critical for information transfer between moving cells.

# 1. Introduction

Intercellular communication via direct cell-cell contact allows the flow of information in tissues during development. Information is exchanged between cells using a diverse set of ligands and receptors expressed on the cells' surfaces, including but not limited to Eph-ephrin, Cadherin-Cadherin, and Delta-Notch receptor-ligand pairs [1-3]. This information is used to coordinate the dynamics of cellular processes across tissues and establish patterns. During development, cellular movement can occur within tissues as they undergo morphogenesis [4,5]. Our interest is how these cellular movements affect the emergence of organized spatial and temporal patterns that yield a reliable developmental program.

The vertebrate segmentation clock provides an attractive model system to address this question, because it involves intercellular signaling [6-9] together with cell movement [10-12]. This clock operates during vertebrate embryonic development, and controls the rhythmic formation of somites, which are the precursors of vertebrae and other tissues that make the characteristic segmented structure of vertebrates. The segmentation clock is a tissue level rhythmic pattern generator consisting of a population of mobile cells in the presomitic mesoderm (PSM) [13,14]. Each cell in the PSM possesses a single-cell genetic oscillator composed by negative feedback loops [15-17]. These oscillators can interact with their neighbors through membrane proteins Delta and Notch, and synchronize their phases locally [6-9].

However, by itself local coupling through Delta-Notch signaling may be a poor way to achieve global phase synchronization across a cell population, as is observed in the tailbud, the tissue at the posterior of the PSM. The reason is that locally coupled oscillators have a strong propensity to form persistent spatial structures that prevent the cell population from reaching global synchronization [18-21]. Recently it has been reported that cells in the posterior PSM move around and exchange their neighbors over time [10-12]. Motivated by these observations, a theoretical study suggested that cell movement observed in the posterior PSM may be important for quickly achieving global synchronization of genetic oscillators across a cell population when cells use a local coupling mechanism [21]. This result is supported by further theoretical studies addressing the effects of mobility on coupled oscillators [18,22-24]. It has been suggested that cells moving faster synchronize their oscillators more quickly by preventing the formation of persistent spatial structures [18].

An important assumption in these previous studies is that when a cell arrives at a new location, it immediately interacts at full capacity with its new neighbors via membrane proteins (such as Delta and Notch; [21]). However, it is reasonable to expect that the association of membrane proteins between two cells that gradually come into contact with each other might need some time to reach full capacity (figure 1(A), see also Materials and methods). Indeed, such a gradual interaction was recently measured for Delta-Notch signaling in a cell culture system [25]. Here we ask whether cellular mobility still improves synchronization when cells have to gradually recover intercellular interactions after movement.

To answer this question, we develop a coupled phase oscillator model in which we take into account both cell movement and the gradual recovery of intercellular interactions after movement. We first derive an equation for the time evolution of the coupling strength between two adjacent cells after they come into contact, by considering the kinetics of membrane protein binding events between these two cells. We use this model to show that the degree of synchronization depends in a non-monotonic way on cell movement. We find (1) an optimal moving rate for synchronization and, (2) a critical moving rate above which cell movement destroys synchronization. We explain these optimal and critical moving rates in terms of the competition of timescales between the moving rate and the coupling recovery rate. Our results indicate that whether the moving rate observed in *vivo* can promote synchronization of genetic oscillators critically depends on the relative speed with which cells establish and develop interactions with their new neighbors after movement.

## 2. Theoretical description of mobile, coupled oscillators

We model a cell population in the tailbud as a discrete system in which cells are located on a two-dimensional lattice. The unit of length is the distance between two adjacent sites in the lattice. Each cell on the lattice is identified by the index $j$ ($j = 1, 2, …, N$). Cells in the bulk can interact with their four nearest neighbors (figure 1(A)), while cells in the boundaries interact with their two or three neighboring cells.

To describe the cell movement in the PSM we allow cells on the lattice to exchange positions with one of their nearest neighbors, at random times with a characteristic time scale $1/\lambda$ (see figure 1(C) and Materials and methods for the procedures of simulations; [21]). The inverse of the characteristic time scale defines the moving rate $\lambda$. For larger $\lambda$, cells exchange their locations more frequently. We assume isotropic cell movement. With this representation of cell movement, a single cell performs a random walk in the lattice, as observed experimentally in the PSM [10,11] and described in [10]. The details of movement processes, for example cell shape changes, have not been experimentally characterized in the PSM yet. Our description of cell movement is a simplification of more complex processes of movement within the tissue that allows several analytical treatments, as shown below. In the boundaries, cells exchange their positions with one of their two or three neighboring cells. This choice of boundary condition for cell movement is motivated by the fact that cells in the PSM are mostly constrained to move within the tissue.

To represent the oscillators, we adopt a locally coupled phase oscillator model [26], as was done in previous theoretical studies on the segmentation clock [21,27,28]. It was shown that the phase oscillator model captures the dynamics of more detailed models that explicitly describe Delta-Notch signaling for the segmentation clock [21]. Moreover, phase oscillators were successfully used to fit theory to experimental data in studies on the segmentation clock [9,29] and the circadian clock [30,31].

We consider the situation in which all cells have identical intrinsic frequency $\omega$ for their intracellular genetic oscillators. For the segmentation clock this intrinsic frequency is determined by the reaction kinetics in negative feedback loops downstream of Delta-Notch signaling [16,32]. To include the dependence of the coupling strength – the interaction rate – on the elapsed time after contact, we introduce a time dependent coupling strength. The phase $\theta_j(t)$ of cell $j$ at time $t$ obeys:

$$\frac{d\theta_j(t)}{dt} = \omega + \frac{1}{n_j} \sum_\alpha \kappa_{j\alpha}(t_{j\alpha}) \sin(\theta_\alpha(t) - \theta_j(t)) + \sqrt{2C} \xi_j(t) \quad \text{for } j = 1, 2, \ldots, N, \tag{1a}$$

where $\sum_\alpha$ represents summation over nearest neighbors, $\kappa_{j\alpha}(t_{j\alpha})$ is the time dependent coupling strength between cell $j$ and cell $\alpha$, $n_j$ is the number of nearest neighbors for cell $j$ ($n_j$ = 2, 3 or 4), $C$ is the noise strength and $\xi_j(t)$ is a white Gaussian noise with $\langle \xi_j(t) \rangle = 0$ and $\langle \xi_j(t) \xi_{j'}(t') \rangle = \delta_{jj'} \delta(t-t')$. To describe the manner in which a cell takes some time to interact at full capacity with its new neighboring cells after movement (figure 1(A)), we introduce the following expression for coupling strength (see Materials and methods for a derivation of this expression where we explicitly consider simple assumptions on the kinetics of binding processes between ligands and receptors on two adjacent cells):

$$\kappa_{j\alpha}(t_{j\alpha}) = \kappa_0 \left(1 - e^{-\beta t_{j\alpha}}\right), \tag{1b}$$

where $\kappa_0$ is the maximum coupling strength, $t_{j\alpha}$ is the elapsed time after cell $j$ and cell $\alpha$ made contact with each other, and $\beta$ is the coupling recovery rate after contact. Larger $\beta$ means a faster recovery of the interaction (figure 1(B)). The coupling strength between two adjacent cells is zero at the moment after they contact with each other. It increases with time as long as these two cells stay adjacent. To focus our analyses on this interaction-recovery process and to simplify the model we assume that a cell that just left its position instantaneously ceases to interact with its old neighbors. The inclusion of a gradual coupling decay between old neighbors in Eq. (1) is an interesting extension of the model that we leave for future work.

To measure the degree of synchronization we use the order parameter proposed by Kuramoto [33]:

$$Z(t) = \left\langle \left| \frac{1}{N} \sum_{j=1}^{N} e^{i\theta_j(t)} \right| \right\rangle_{ens}, \qquad (2)$$

where $i = \sqrt{-1}$ and $\langle ... \rangle_{ens}$ represents the average over the different realizations of initial conditions and noise (for initial conditions, see Materials and methods). If $Z(t)$ is close to unity, the phases of oscillators are relatively close to each other and the system is in a synchronized state. In contrast, if $Z(t)$ is close to zero, phases are scattered and the system is in an unsynchronized state.

Although the order parameter Eq. (2) measures the degree of synchronization, it cannot characterize phase profiles that emerge in a model that includes space [18]. In order to characterize spatial phase profiles and to measure the local phase order, we also consider the correlation between two sites at distance $d$ in the two-dimensional lattice:

$$\rho(d,t) = \langle \cos(\vartheta_\mathbf{k}(t) - \vartheta_{\mathbf{k'}}(t)) \rangle_{|\mathbf{k}-\mathbf{k'}|=d}, \qquad (3)$$

where $\mathbf{k} = (k, l)$ represents a site in the two-dimensional lattice, $\vartheta_\mathbf{k}(t)$ represents the phase value of site $\mathbf{k}$ at time $t$ (e.g. if cell $j$ is in site $\mathbf{k}$ at time $t$, $\vartheta_\mathbf{k}(t) = \theta_j(t)$) and $\langle ... \rangle_{|\mathbf{k}-\mathbf{k'}|=d}$ represents an average over all pairs of sites between which the distance is $d$. The value of $\rho$ lies between $-1$ and $1$. If two sites at a distance $d$ of each other tend to have similar phase value, $\rho$ is close to 1. In contrast if they tend to be opposite in phase, $\rho$ is close to $-1$. If there is no correlation between them, $\rho \approx 0$. To get better statistics we calculated an average of $\rho$ over different realizations of initial conditions and noise.

The values of parameters in the model for simulations are listed in Table 1. In Materials and methods we estimate the moving rate $\lambda$ of PSM cells as roughly around $0.05 \sim 0.1$ min$^{-1}$, from the data in previous studies on chick somitogenesis [10,11]. Below, we explore a wide range of the moving rate $\lambda$ including these estimated values.

## 3. Optimal and critical moving rates for synchronization

To study the effect of gradual coupling recovery we numerically simulate Eq. (1) and measure the degree of synchronization by Eqs. (2) and (3). Figures. 2(*A*) and (*B*) show the snapshots of phase profiles and the time evolution of the order parameter $Z(t)$ defined by Eq. (2), respectively, for a fixed coupling recovery rate $\beta = 33\kappa_0$ (hereafter we use $\kappa_0 = 0.03$ min$^{-1}$ as the unit of time, see Materials and methods). When cells do not move ($\lambda = 0$, red triangles in figure 2(*B*)), the order parameter increases with time and finally approaches a steady state value around $Z = 0.5$. The standard deviations of the order parameter $Z(t)$ are large because a single trajectory of the order parameter fluctuates strongly due to noise, and the time needed for the trajectory to reach the steady state value is sensitive to the initial phase differences (figure S1). As indicated by the sharp decrease of $\rho$ with increasing distance $d$ in figure 2(*C*), these non-mobile cells can reach and maintain short-range correlation of phases, but they cannot achieve long-range correlation even after a long time (see also figure 2(*A*)). This means that these cells tend to form local clusters of synchronization between which phases differ greatly. When cells exchange their locations every $1/4\kappa_0$ on average ($\lambda = 4\kappa_0$, green circles in figure 2(*B*)), these cells build synchronization more rapidly compared to when they do not move. Moreover, the steady state value of the order parameter is much larger, around $Z = 0.7$. In this case we observe both short- and long-range correlations of phases (figures 2(*A*) and (*C*)), confirming that these cells achieve global phase order. This result is consistent with previous work using a gene network model, which shows that cell movement promotes synchronization for $\beta \gg \lambda$ (see supplementary figure S6 in [21]).

However, if each cell exchanges its position more frequently than considered above ($\lambda = 40\kappa_0$, blue squares in figure 2(*B*)), the degree of synchronization is much worse than when the cells do not move ($\lambda = 0$). In this case we observe phase disordered states where neither local nor global phase order exists, as represented by $\rho \approx 0$ for any $d$ in figure 2(*C*) (see also figure. 2(*A*)). This indicates that the dependence of the degree of synchronization on the moving rate is non-monotonic, suggesting the existence of an optimal moving rate to achieve synchronization. We systematically studied the behavior of the order parameter for a range of moving rates, confirming the existence of an optimal moving rate (figure 2(*D*)). In addition, we found a transition point in the moving rate: if the moving rate is smaller than a critical moving rate $\lambda^*$, global phase order appears, while if it is larger than $\lambda^*$ the system goes to phase disorder (figure 2(*D*), $\lambda \approx 23.3\kappa_0$). Thus, cells have to move at an appropriate rate to achieve better synchronization when they need to recover the interactions with their new neighboring cells after movement.

Both the optimal and critical moving rates depend on the coupling recovery rate $\beta$ (figure 3(A)). When $\beta$ is small ($\beta = 3.3\kappa_0$, red filled triangles in figure 3(A)), no apparent peak is observable, meaning that cell movement cannot improve the degree of synchronization across the population of oscillators. In contrast, with the increase of $\beta$, an optimal moving rate for synchronization appears ($\beta = 10.5\kappa_0 \sim 333\kappa_0$ in figure 3(A)) and there is a range of the moving rate for which the degree of synchronization is better than that of non-mobile oscillators. As $\beta$ increases, this range becomes wider and both the optimal and critical moving rates become larger. These results indicate that whether cell movement at a given moving rate promotes the synchronization of oscillators, depends on the coupling recovery rate. Note that the degree of synchronization achieved at each moving rate also increases with the increase in $\beta$, and eventually saturates to the value corresponding to instantaneous coupling recovery (*i.e.* for $\kappa_{j\alpha}(t_{j\alpha}) \equiv \kappa_0$ in Eq. (1a), black solid line in figure 3(A)). For instantaneous coupling recovery, the order parameter $Z$ monotonically increases with the increasing moving rate $\lambda$, as was previously shown by [21].

To expose the competition of timescales occurring between the moving rate $\lambda$ and the coupling recovery rate $\beta$, we examine how the degree of synchronization in figure 3(A) scales with the ratio $\lambda/\beta$. The collapse of the curves is not complete, showing that the order parameter $Z$ is not a single function of this ratio, but rather it depends on $\lambda$ and $\beta$ independently (figure 3(B)). However, the transition point from phase order to phase disorder coincides at around $\lambda/\beta = 0.7$ for large enough $\beta$. This implies that the transition point does not depend on individual values of $\lambda$ and $\beta$, but on the ratio $\lambda/\beta$ if $\beta$ is large.

We next examined how noise affects optimal and critical moving rates. As the noise strength $C$ increases, the critical moving rate becomes smaller (figure 4(A)). This result indicates that noise reduces the range of $\lambda/\beta$ in which cell movement enhances the degree of synchronization. When $C$ is small, the maximum degree of synchronization at the optimal moving rate is not very pronounced in steady states (*e.g.* $C = 0.1\kappa_0$ in figure 4(A)). However, even for small $C$, there is an optimal moving rate at which cells synchronize much faster than non-mobile cells (figure 2(D)). This optimal moving rate becomes smaller as $C$ increases. In summary, both the coupling recovery rate $\beta$ and the noise strength $C$ determine the range in which cell movement can improve the degree of synchronization.

## 4. The origin of optimal and critical moving rates

To understand the optimal moving rate and to estimate the critical moving rate, we introduce an effective coupling strength. This effective coupling strength approximates the time varying coupling strength between each pair of adjacent sites in the two-dimensional lattice (figure 5) by its temporal average.

Let $\kappa_{\mathbf{kk'}}(t)$ be the time series of the coupling strength between a pair of adjacent *sites*, $\mathbf{k}$ and $\mathbf{k'}$, in the two-dimensional lattice (for example $\mathbf{k} = (k, l)$ and $\mathbf{k'} = (k, l+1)$, figures 5(*A*) and S2). As long as a pair of cells stays adjacent to each other in these two sites, $\kappa_{\mathbf{kk'}}(t)$ increases with time according to Eq. (1b). This interaction time ends when one of these two cells moves away from these two sites, and $\kappa_{\mathbf{kk'}}(t)$ is reset to zero. Note that the length of the interaction time is stochastic due to random cellular motions.

The time average of $\kappa_{\mathbf{kk'}}(t)$ between $t_0$ and $t_0 + T$ is defined as:

$$\langle \kappa \rangle_T = \frac{1}{T} \int_{t_0}^{t_0+T} \kappa_{\mathbf{kk'}}(t) dt, \qquad (4)$$

where *T* is the time window for averaging. We can calculate this time average analytically by using Eq. (1b) and the probability density function for the length of the interaction time, assuming that *T* is sufficiently large (see Supporting Information for detailed calculation). The time average of the coupling strength between a pair of adjacent sites in the bulk of the two-dimensional lattice can be written as:

$$\langle \kappa \rangle_T = \frac{1}{1 + 3\lambda/2\beta} \kappa_0. \qquad (5)$$

Eq. (5) shows that the time average of coupling strength $\langle \kappa \rangle_T$ is a decreasing function of $\lambda/\beta$, the ratio of the moving rate $\lambda$ to the coupling recovery rate $\beta$ (figure 5(*B*)). Moving faster and/or recovering the interactions slower, reduces the effective coupling strength between neighboring cells. If $\lambda/\beta << 1$ then $\langle \kappa \rangle_T \approx \kappa_0(1 - 3\lambda/2\beta)$. In contrast, if coupling recovery is very slow and/or cells move very fast (*i.e.* $\lambda/\beta >> 1$), $\langle \kappa \rangle_T \approx 2\kappa_0\beta/3\lambda$.

Eq. (5) agrees with the numerically calculated time average of the coupling strength between two adjacent sites in the two-dimensional lattice (figure S2(*C*) for the time window *T* = $300\kappa_0^{-1}$). Furthermore, the time evolution of the order parameter calculated from the simulations of Eq. (1a) with $\kappa_{j\alpha}(t_{j\alpha}) \equiv \langle \kappa \rangle_T$ approximates that of the original model Eq. (1a) with Eq. (1b) very well (colored solid lines in figures 2, 3 and 4).

Using Eq. (5) we can explain why a critical moving rate for synchronization appears (figure 5(*B*)). For coupled noisy phase oscillators it is known that there is a critical coupling strength $\kappa^*$ below which the incoherent (phase disordered) state becomes stable and global phase order is not possible [18,34]. If the ratio $\lambda/\beta$ is sufficiently large, $\langle\kappa\rangle_T$ becomes smaller than $\kappa^*$. Hence, cells moving faster and/or recovering interaction slower cannot achieve synchronization.

Moreover, Eq. (5) provides a heuristic interpretation for the dependence of the critical moving rate on the noise strength $C$. The noise strength $C$ determines the critical coupling strength $\kappa^*$, if $C$ is large then $\kappa^*$ is also large. Therefore, the effective coupling strength $\langle\kappa\rangle_T$ becomes smaller than $\kappa^*$ even at smaller $\lambda/\beta$. Hence, synchronization is allowed only within a reduced interval of $\lambda/\beta$ if the noise is strong. In contrast, if the noise is weak, $\kappa^*$ is smaller. Therefore, the interval in which $\langle\kappa\rangle_T$ is larger than $\kappa^*$ becomes wider, allowing cells moving relatively faster to realize synchronization (figure 4).

Finally, the emergence of an optimal moving rate is explained as follows: If $\lambda/\beta$ is sufficiently small, $\langle\kappa\rangle_T$ can be larger than $\kappa^*$ (figure 5(B)). However, if cells move slowly enough, spatial structures that tend to hamper the achievement of global synchronization are more likely to appear (see $\lambda = 0$ in figure 2(A) and (C)). Hence an optimal value of $\lambda/\beta$ exists where $\langle\kappa\rangle_T$ is larger than $\kappa^*$, yet cells move fast enough to prevent the formation of persistent spatial structures (figures 3 and 5).

## 5. Estimation of the critical moving rate

The critical moving rate $\lambda^*$ above which synchronization breaks down can be obtained from an expression for the critical coupling strength $\kappa^*$, and the expression for the effective coupling strength $\langle\kappa\rangle_T$, Eq. (5) (figure 5(B)). Therefore, to calculate $\lambda^*$ we first need to know the critical coupling strength $\kappa^*$ of mobile oscillators in a two-dimensional lattice.

Consider cells moving fast enough to meet all the cells in the two-dimensional lattice in a sufficiently short time. In such a situation, we speculate that their behavior can be approximated by that of oscillators with mean field coupling (*i.e.* all-to-all coupling; [22-24]). For oscillators with mean field coupling, the critical coupling strength $\kappa_m^*$ below which the incoherent state is stable is:

$$\kappa_m^* = 2C_m, \tag{6}$$

where $C_m$ is the noise strength in the mean field system [34]. We approximate the critical coupling strength of mobile oscillators in a two-dimensional lattice as $\kappa^* \approx 2C$, as long as these oscillators move very fast. Therefore, the intersection between Eq. (5) and $\kappa^*$ gives:

$$\frac{1}{1+3\lambda^*/2\beta}\kappa_0 = 2C \:. \tag{7}$$

From Eq. (7) we can obtain the critical moving rate as:

$$\frac{\lambda^*}{\beta} = \frac{\kappa_0/C - 2}{3} \:. \tag{8}$$

Vertical lines in figures 3(*B*) and 4(*A*) indicate $\lambda^*/\beta$ calculated using Eq. (8) with the corresponding values of $\kappa_0$ and *C* used in simulations, showing that Eq. (8) is a good approximation for the transition point from phase order to phase disorder. Eq. (8) further explains the scaling of the transition point among different values of *β* observed in figure 3(*B*) as *β* becomes large, for a fixed noise strength $C = 0.25\kappa_0$. Introducing the rescaled variable $\tilde{\lambda} = 3(\lambda/\beta)/(\kappa_0/C - 2)$, Eq. (8) predicts that the transition point should be at $\tilde{\lambda} = 1$. This is in fact what we see in figure 4(*B*), where the collapse of curves with different noise strengths occurs close to this transition point.

## 6. Discussion

In this paper we developed a mathematical model that includes the gradual recovery of signaling between newly contacted cells after movement (figure 1), and used this model to study how this gradual recovery affects synchronization of genetic oscillators. We showed the existence of an optimal moving rate for synchronization and a critical moving rate above which synchronization is not possible (figure 2). These optimal and critical moving rates depend on the coupling recovery rate and the noise strength (figures 3 and 4). By considering the time average of the time-varying coupling strength, we derived an expression for an effective coupling strength, Eq. (5). This effective coupling strength reveals how the optimal and critical moving rates emerge in the system (figure 5). These interesting new properties are a direct consequence of the gradual recovery of intercellular interactions after movement and do not occur with instantaneous recovery [21]. We confirmed that our conclusions still hold even in the presence of time delays in intercellular communication [27,29] (see Supporting Information and figures S3 and S4). Here, to highlight the effects of the coupling recovery and make the analysis of Eq. (1a) simpler, we neglected coupling time delays.

Our results suggest that by controlling the moving rate or the recovery rate, the speed and degree of synchronization of a system can be altered experimentally. Recently, some key parameters involved in an effective theoretical description of the segmentation clock have been estimated by fitting theory to experimental data, including the coupling strength between cells, and noise strength [9,29]. This system may offer several ways to test the effect of movement on synchronization. A naturally occurring gradient in cellular mobility has been observed along the PSM [10,12] and different synchronization speeds may be found at different locations in the mobility gradient after experimental desynchronization. It may be possible to experimentally alter the kinetics of Delta-Notch signaling in the PSM, for example by changing the affinity of ligand and receptor [35] and thereby alter the coupling recovery rate. Moreover, a chemical factor (fibroblast growth factor) that affects the mobility of these cells has also been identified [10,11]. These studies suggest the possibility to regulate the ratio of the moving and recovery rate experimentally and to test whether an optimal and a critical mobility exist as the current theoretical study predicts.

The finding of synchronization optima also raises the possibility that a synchronized multicellular biological clock may have evolved to an optimal ratio of the moving and recovery rate. For a system sitting in this optimal ratio, our study predicts that any perturbation to mobility or coupling recovery would be detrimental to synchronization. Furthermore, by increasing this ratio beyond the critical value, synchronization would be severely disrupted.

In other developmental processes like epiboly, gastrulation, and convergence-extension, as well as in pathological situations such as inflammation and metastasis, cell movements take place at the same time as cell-cell signaling [4,5,36-39]. In microbial communities and in pathogen-host interactions, cell-contact signaling and movement also occur together [40]. In these processes, the type of cellular movement might differ from that observed in the PSM. However, the essential point is that cellular mobility can cause a new contact between cells that were previously distant from each other. Generally, in cellular systems intercellular communication is established and developed gradually over a characteristic time. Our study indicates that the competition between this characteristic time scale and that given by mobility is key to the ability of the system to organize spatiotemporal patterns.

**7. Materials and methods**
*7.1 Coupling strength dynamics*
To derive the coupling strength given by Eq. (1b) we consider an event in which two adjacent cells begin to interact with each other through receptors and ligands expressed on their cell surfaces. We make a few simplifying assumptions about the kinetics of binding events to obtain an expression for the resulting coupling strength. Let $p(t)$ be the amount of receptor-ligand pairs already bound at time $t$. The time evolution of $p(t)$ is given by the following equation:

$$\frac{dp}{dt} = \beta(p_s - p), \tag{9}$$

where $p_s$ is the saturating capacity (*i.e.* the total amount of receptor-ligand pairs available per cell-cell contact) and $\beta$ is the rate of binding processes. The difference $p_s - p$ is the amount of receptor-ligand pairs that have not yet bound together. In reality, $p_s$ might change with time due to the increase or decrease of the total amount of receptors and ligands. However, here we assume for simplicity that $p_s$ is constant over time. With this assumption, Eq. (9) can be solved analytically with the initial condition $p(0) = 0$:

$$p(t)/p_s = 1 - e^{-\beta t}. \tag{10}$$

Assuming that the coupling strength is proportional to the ratio $p(t)/p_s$ gives us Eq. (1b).

In the segmentation clock, the coupling strength $\kappa$ is related to the number of available receptor molecules, Notch and ligand Delta [6-9]. Apart from the Delta-Notch signaling pathway, no other mechanism has been identified in this system to mediate intercellular coupling.

In a tissue, moving cells come into contact gradually, and consequently their contact surface grows gradually [36]. Our model for cell movement does not include such gradual contact growth, because cells in our model just exchange their positions in a two-dimensional lattice when they move. Thus, Eq. (1b) can be interpreted as an effective description for gradual contact growth in the context of a lattice model.

*7.2 The integration scheme for Eq. (1)*

In this paper we use a two-dimensional square lattice. Cells in the lattice exchange positions with their neighbors at random times. We assume that the probability that each cell changes its location per unit time is $\lambda$. The time interval until the next exchange event occurs is determined using the Gillespie algorithm [41]. Eq. (1) is then integrated with the Heun method between two successive exchange events. The time step $\Delta t_s$ for numerical integration is fixed during a single simulation, and is determined such that for a given value of the moving rate $\lambda$, $\Delta t_s$ is smaller than the average time interval until the next exchange event $1/(\lambda N/2)$, as $10\Delta t_s \approx 1/(\lambda N/2)$. If $\Delta t_s > 0.01$ in this equation, we set $\Delta t_s = 0.01$. In the simulations, when a time interval until the next exchange event given by the Gillespie algorithm is smaller than $\Delta t_s$, we accept this exchange event at the next time point.

*7.3 Parameter values*

Throughout this paper we fix $\omega = 0$ without loss of generality, considering the system from a co-rotating reference frame. We use $\kappa_0 = 0.03$ min$^{-1}$, which is based on estimations of the coupling strength in vertebrate somitogenesis obtained from theory fits to experimental data [9,29]. However, our results are not qualitatively sensitive to the value of $\kappa_0$. Similarly we take a typical size of the relevant tissue during somitogenesis, a square lattice of $16 \times 16$, with a total of $N = 256$ cells [21]. As the domain size becomes larger, oscillators need more time to achieve synchronization (figure S5). In addition, a steady state value of the order parameter $Z$ is smaller for a larger domain size than for a smaller one. However, for different domain sizes our main results are qualitatively the same (figure S5). The choice of boundary conditions does not affect our conclusions either (figure S6).

Recent measurements of cellular mobility in the posterior chick PSM [10,11] provided values of cellular velocity of about 0.5~1.0 μm/min. Assuming the average cell diameter of 10 μm, these data suggest that cells move about one cell diameter roughly in 10 ~ 20 min, which gives the moving rate of 0.05 ~ 0.1 min$^{-1}$. This observed cell movement seems rapid enough compared to the period of the chick segmentation clock, which is around 90 min [42]. In fact, a previous study demonstrated that cell movement at the rates estimated above can improve synchronization even under an oscillatory period of 30 min in simulations [21].

*7.4 Initial conditions*

The initial phase of each oscillator is chosen randomly from a uniform distribution between 0 and $2\pi$. In this paper the order parameter Eq. (2) and the correlation Eq. (3) are calculated from 200 realizations of the initial conditions and noise unless otherwise indicated. At the initial time, the coupling strength for each pair of adjacent cells is set to its maximum capacity $\kappa_0$. If a cell contacts with new cells after movement in the simulations, the coupling strength between them changes according to Eq. (1b).

*7.5 Steady state measurements*

The simulations are run long enough for the order parameter defined by Eq. (2) and the correlation defined by Eq. (3) to reach steady state values (figure S7). The time taken to reach steady state values strongly depends on the moving rate $\lambda$. For this reason we use different calculation times, ranging from $150\kappa_0^{-1}$ to $1500\kappa_0^{-1}$ depending on the value of $\lambda$, to optimize computational costs. The steady state measurement we report in figures 2-4 is the temporal average of the order parameter and the correlation calculated using the last $90\kappa_0^{-1}$ time span.

**Acknowledgements**

We thank the group of Prof. Jülicher and the Max Planck Institute for the Physics of Complex Systems for their hospitality and insightful discussions. KU is supported by the Japan Society for the Promotion of Science for Young Scientists. SA acknowledges grant MOSAICO (Ministerio de Ciencia e Innovación, Spain). LGM acknowledges CONICET and ANPCyT PICT 876. ACO and LGM are supported by the Max Planck Society and by the European Research Council under the European Communities Seventh Framework Programme (FP7/2007-2013) / ERC Grant no. 207634.

**References**

[1]     Andersson E R, Sandberg R and Lendahl U 2011 Notch signaling: simplicity in design, versatility in function *Development* **138** 3593-612

[2]     Pitulescu M E and Adams R H 2010 Eph/ephrin molecules-a hub for signaling and endocytosis *Gene Dev* **24** 2480-92

[3]     Stepniak E, Radice G L and Vasioukhin V 2009 Adhesive and signaling functions of cadherins and catenins in vertebrate development *Cold Spring Harb Perspect Biol* **1** a002949

[4]     Friedl P and Gilmour D 2009 Collective cell migration in morphogenesis, regeneration and cancer *Nat Rev Mol Cell Biol* **10** 445-57

[5]     Solnica-Krezel L 2005 Conserved patterns of cell movements during vertebrate gastrulation *Curr Biol* **15** R213-28

[6]     Horikawa K, Ishimatsu K, Yoshimoto E, Kondo S and Takeda H 2006 Noise-resistant and synchronized oscillation of the segmentation clock *Nature* **441** 719-23

[7]     Jiang Y J, Aerne B L, Smithers L, Haddon C, Ish-Horowicz D and Lewis J 2000 Notch signalling and the synchronization of the somite segmentation clock *Nature* **408** 475-9

[8]     Özbudak E M and Lewis J 2008 Notch signalling synchronizes the zebrafish segmentation clock but is not needed to create somite boundaries *PLoS Genet* **4** e15

[9]     Riedel-Kruse I H, Müller C and Oates A C 2007 Synchrony dynamics during initiation, failure, and rescue of the segmentation clock *Science* **317** 1911-5

[10]    Bénazéraf B, Francois P, Baker R E, Denans N, Little C D and Pourquié O 2010 A random cell motility gradient downstream of FGF controls elongation of an amniote embryo *Nature* **466** 248-52

[11]    Delfini M C, Dubrulle J, Malapert P, Chal J and Pourquié O 2005 Control of the segmentation process by graded MAPK/ERK activation in the chick embryo *Proc Natl Acad Sci USA* **102** 11343-8

[12]    Mara A, Schroeder J, Chalouni C and Holley S A 2007 Priming, initiation and synchronization of the segmentation clock by deltaD and deltaC *Nat Cell Biol* **9** 523-30

[13]    Oates A C, Morelli L G and Ares S 2012 Patterning embryos with oscillations: structure, function and dynamics of the vertebrate segmentation clock *Development* **139** 625-39

[14]    Pourquié O 2011 Vertebrate segmentation: from cyclic gene networks to scoliosis *Cell*

**145** 650-63

[15] Hirata H, Yoshiura S, Ohtsuka T, Bessho Y, Harada T, Yoshikawa K and Kageyama R 2002 Oscillatory expression of the bHLH factor Hes1 regulated by a negative feedback loop *Science* **298** 840-3

[16] Lewis J 2003 Autoinhibition with transcriptional delay: A simple mechanism for the zebrafish somitogenesis oscillator *Curr Biol* **13** 1398-408

[17] Oates A C and Ho R K 2002 Hairy/E(spl)-related (Her) genes are central components of the segmentation oscillator and display redundancy with the Delta/Notch signaling pathway in the formation of anterior segmental boundaries in the zebrafish *Development* **129** 2929-46

[18] Peruani F, Nicola E M and Morelli L G 2010 Mobility induces global synchronization of oscillators in periodic extended systems *New J Phys* **12** 093029

[19] Tiedemann H B, Schneltzer E, Zeiser S, Rubio-Aliaga I, Wurst W, Beckers J, Przemeck G K H and De Angelis M H 2007 Cell-based simulation of dynamic expression patterns in the presomitic mesoderm *J Theor Biol* **248** 120-9

[20] Tinsley M R, Taylor A F, Huang Z Y and Showalter K 2009 Emergence of collective behavior in groups of excitable catalyst-loaded particles: Spatiotemporal dynamical quorum sensing *Phys Rev Lett* **102** 158301

[21] Uriu K, Morishita Y and Iwasa Y 2010 Random cell movement promotes synchronization of the segmentation clock *Proc Natl Acad Sci U S A* **107** 4979-84

[22] Frasca M, Buscarino A, Rizzo A, Fortuna L and Boccaletti S 2008 Synchronization of moving chaotic agents *Phys Rev Lett* **100** 044102

[23] Fujiwara N, Kurths J and Díaz-Guilera A 2011 Synchronization in networks of mobile oscillators *Phys Rev E* **83** 025101

[24] Skufca J D and Bollt E M 2004 Communication and synchronization in disconnected networks with dynamic topology: Moving neighborhood networks *Math Biosci Eng* **1** 347-59

[25] Sprinzak D, Lakhanpal A, Lebon L, Santat L A, Fontes M E, Anderson G A, Garcia-Ojalvo J and Elowitz M B 2010 Cis-interactions between Notch and Delta generate mutually exclusive signalling states *Nature* **465** 86-90

[26] Sakaguchi H, Shinomoto S and Kuramoto Y 1987 Local and global self-entrainments in oscillator lattices *Prog Theor Phys* **77** 1005-10

[27] Morelli L G, Ares S, Herrgen L, Schröter C, Jülicher F and Oates A C 2009 Delayed

coupling theory of vertebrate segmentation *HFSP J* **3** 55-66

[28] Murray P J, Maini P K and Baker R E 2011 The clock and wavefront model revisited *J Theor Biol* **283** 227-38

[29] Herrgen L, Ares S, Morelli L G, Schröter C, Jülicher F and Oates A C 2010 Intercellular coupling regulates the period of the segmentation clock *Curr Biol* **20** 1244-53

[30] Liu C, Weaver D R, Strogatz S H and Reppert S M 1997 Cellular construction of a circadian clock: Period determination in the suprachiasmatic nuclei *Cell* **91** 855-60

[31] Yang Q, Pando B F, Dong G G, Golden S S and Van Oudenaarden A 2010 Circadian gating of the cell cycle revealed in single cyanobacterial cells *Science* **327** 1522-6

[32] Uriu K, Morishita Y and Iwasa Y 2009 Traveling wave formation in vertebrate segmentation *J Theor Biol* **257** 385-96

[33] Kuramoto Y 1984 *Chemical Oscillations, Waves, and Turbulence* (Berlin, Springer-Verlag).

[34] Strogatz S H and Mirollo R E 1991 Stability of incoherence in a population of coupled oscillators *J Stat Phys* **63** 613-35

[35] Cordle J, Redfield C, Stacey M, Van Der Merwe P A, Willis A C, Champion B R, Hambleton S and Handford P A 2008 Localization of the delta-like-1-binding site in human notch-1 and its modulation by calcium affinity *J Biol Chem* **283** 11785-93

[36] Bertet C, Sulak L and Lecuit T 2004 Myosin-dependent junction remodelling controls planar cell intercalation and axis elongation *Nature* **429** 667-71

[37] Keller R 2002 Shaping the vertebrate body plan by polarized embryonic cell movements *Science* **298** 1950-4

[38] Rorth P 2011 Whence directionality: Guidance mechanisms in solitary and collective cell migration *Dev Cell* **20** 9-18

[39] Wallingford J B, Fraser S E and Harland R M 2002 Convergent extension: The molecular control of polarized cell movement during embryonic development *Dev Cell* **2** 695-706

[40] Bassler B L and Losick R 2006 Bacterially speaking *Cell* **125** 237-46

[41] Gillespie D T 1977 Exact stochastic simulation of coupled chemical-reactions *J Phys Chem* **81** 2340-61

[42] Palmeirim I, Henrique D, Ish-Horowicz D and Pourquié O 1997 Avian hairy gene expression identifies a molecular clock linked to vertebrate segmentation and somitogenesis *Cell* **91** 639-48

**Table 1.** Parameters in the model.

| parameters | | values used in simulations |
|---|---|---|
| $\omega$ | intrinsic frequency | 0 |
| $\kappa_0$ | maximum coupling strength | 0.03 min$^{-1}$ |
| $\beta$ | coupling recovery rate | $3.3\kappa_0$ to $333\kappa_0$ |
| $\lambda$ | moving rate | 0 to $300\kappa_0$ |
| $C$ | noise strength | $0.1\kappa_0$ to $0.3\kappa_0$ |
| $N$ | number of cells in the system | 256 (16×16) |

See also Materials and methods for these choices of parameter values.

**Figures**

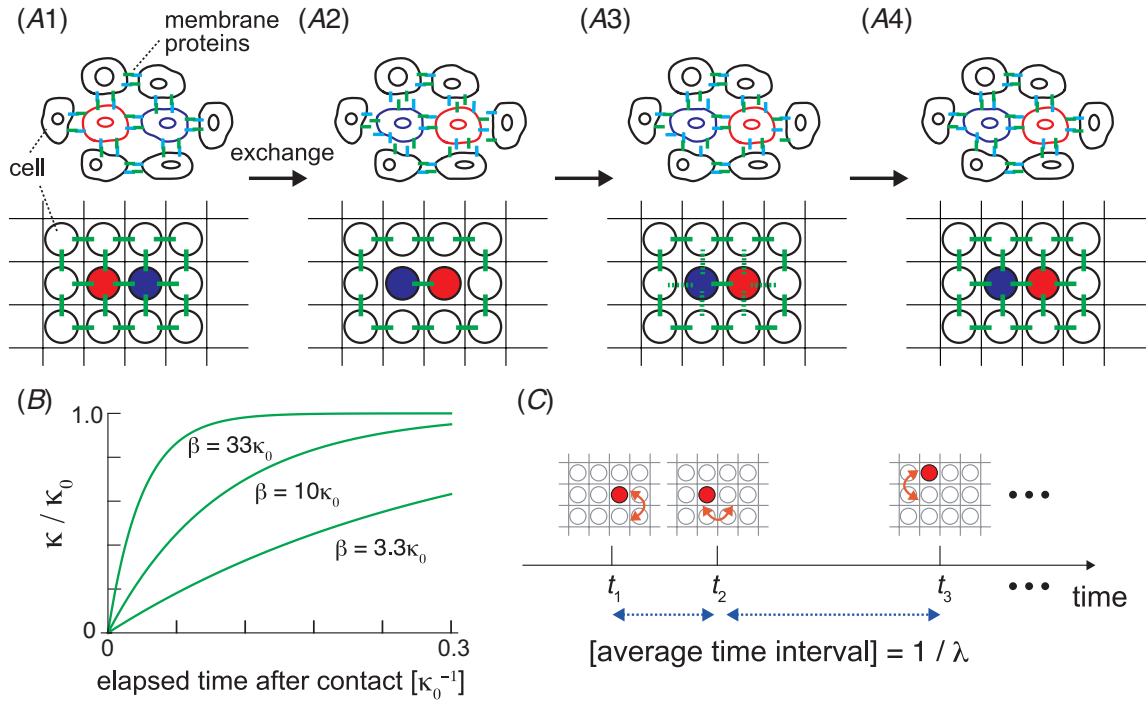

**Figure 1.** Cells need to recover interactions with their new neighbors after movement. (A1) Cells in a two-dimensional lattice. Two cells connected with a solid green line can interact with each other. (A2) The red and blue cells exchanged their locations. They cannot interact with their new neighbors instantaneously. Note that the red and blue cells maintain contact and interact with each other. (A3) and (A4) Interactions are gradually recovered. (B) Time evolution of the coupling strength defined by Eq. (1b). (C) Each cell experiences an exchange of its location every $1/\lambda$ time unit on average. The exchange times obey Poissonian statistics.

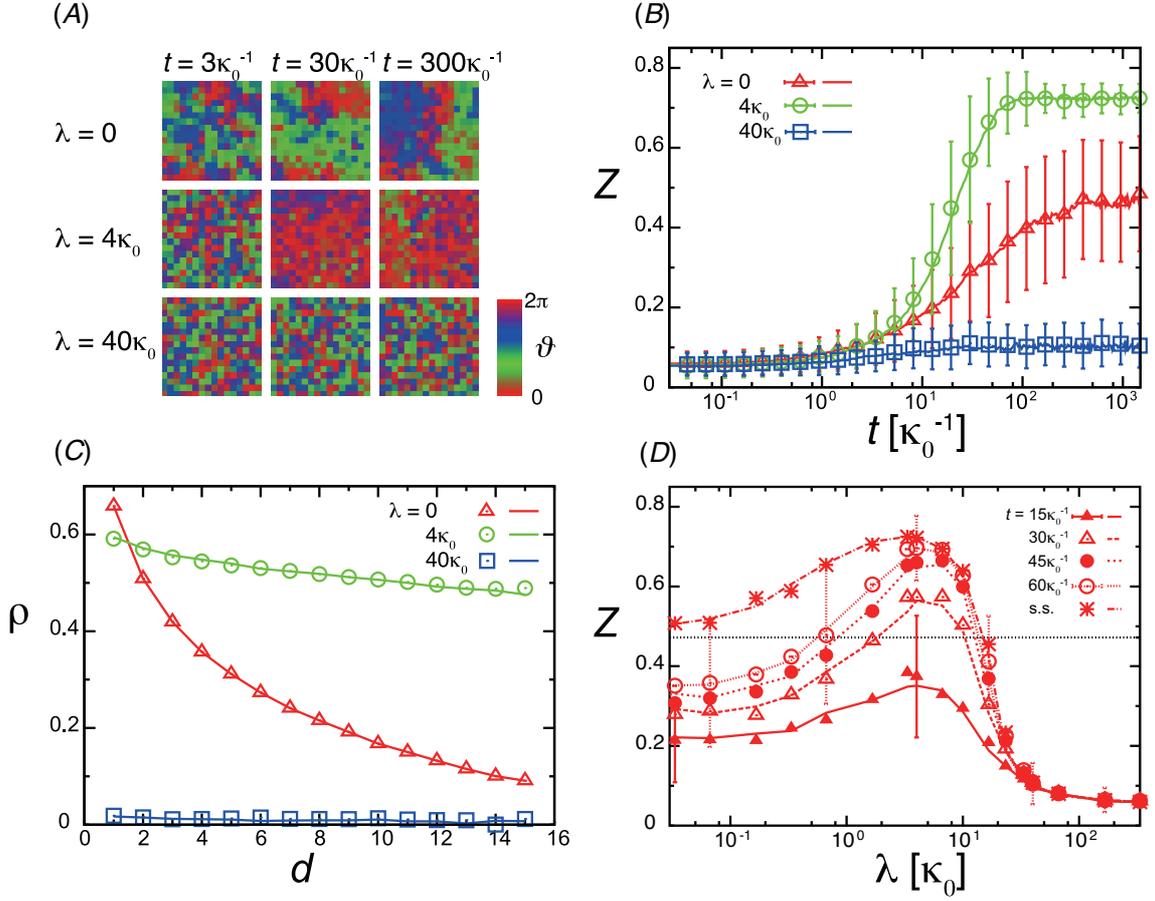

**Figure 2.** Dependence of the degree of synchronization on the moving rate $\lambda$ is non-monotonic. (A) Snapshots of spatial phase profiles in the two-dimensional lattice at different moving rates $\lambda$ observed in numerical simulations of Eq. (1a) with coupling recovery Eq. (1b). The phase $\vartheta$ at each site is represented by a color look-up table as indicated. In (B)-(D) the symbols indicate the results of simulations of Eq. (1a) with coupling recovery Eq. (1b), while the lines indicate the results of simulations of Eq. (1a) with the effective coupling strength Eq. (5). (B) Time evolution of the order parameter $Z(t)$ defined by Eq. (2) at different moving rates $\lambda$ as indicated. Error bars indicate standard deviations (SD). (C) Dependence of the correlation defined by Eq. (3) on the distance between two sites. We plotted the temporal average of the correlation after its value reached a steady state. Error bars for the temporal SD of correlation are smaller than the size of symbols. (D) Dependence of $Z$ on the moving rate $\lambda$. Different symbols and lines correspond to different time points as indicated. s.s. is the steady state. The black horizontal line indicates the steady state value of $Z$ when $\lambda = 0$. Error bars indicate SD. In all panels $\beta = 33\kappa_0$ and $C = 0.25\kappa_0$.

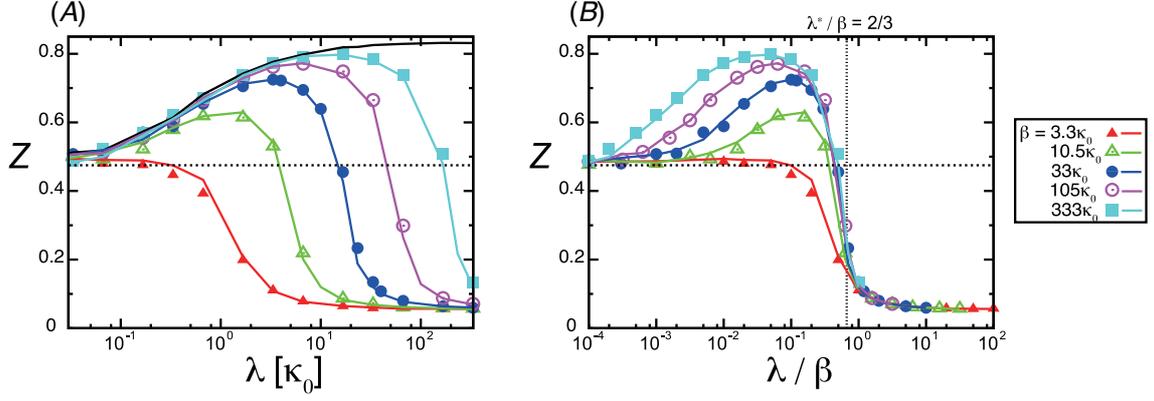

**Figure 3.** The optimal and critical moving rates for synchronization depend on the coupling recovery rate $\beta$. (A) Dependence of the order parameter $Z$ defined by Eq. (2) on $\beta$ at each moving rate $\lambda$. The black solid line shows the order parameter when cells can instantaneously recover interactions with their new neighbors after movement (*i.e.* $\kappa_{j\alpha}(t_{j\alpha}) \equiv \kappa_0$ in Eq. (1a)). (B) Dependence of the order parameter $Z$ defined by Eq. (2) on $\beta$ at each ratio $\lambda/\beta$. The vertical dotted line indicates the transition point $\lambda^*/\beta$ calculated from Eq. (8). In both panels the symbols indicate the results of simulations of Eq. (1a) with coupling recovery Eq. (1b), while the solid lines indicate the results of simulations of Eq. (1a) with the effective coupling strength Eq. (5). The horizontal dotted lines in both panels indicate the steady state value of the order parameter when $\lambda = 0$. In both panels, $C = 0.25\kappa_0$. We ran the simulations long enough for the order parameter $Z(t)$ to reach its steady state value. We calculated the time averages of $Z$ as described in Materials and methods. Error bars for the temporal standard deviations of $Z$ are smaller than the size of symbols.

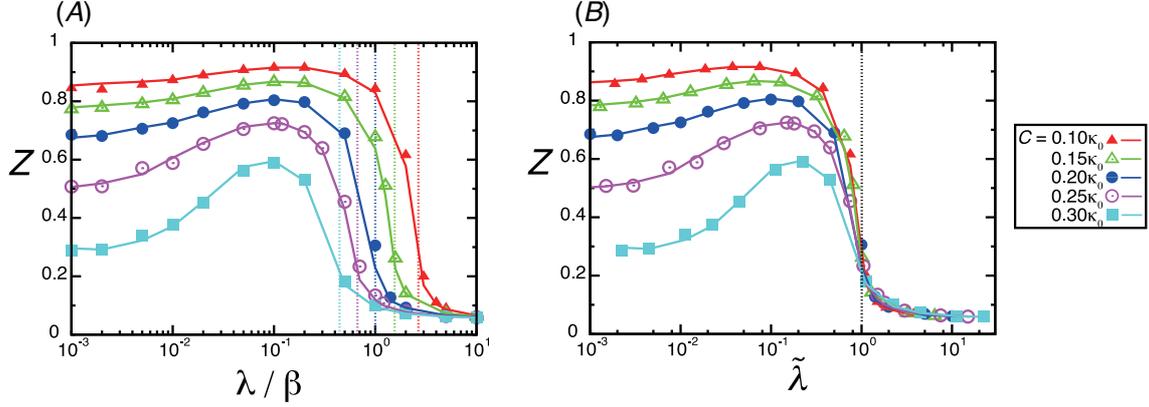

**Figure 4.** Noise limits the range where cell movement can improve synchronization. (A) Order parameter $Z$ vs. the ratio $\lambda/\beta$ of the moving rate $\lambda$ to the coupling recovery rate $\beta$ for different values of the noise strength $C$. Each colored vertical line indicates $\lambda^*/\beta$ calculated from Eq. (8) for the corresponding value of $C$, indicated by the line color. (B) Dependence of order parameter $Z$ on the scaling parameter $\tilde{\lambda} = 3(\lambda/\beta)/(\kappa_0/C - 2)$. The vertical line at $\tilde{\lambda} = 1$ indicates the transition point from phase order to phase disorder calculated from Eq. (8). In both panels the symbols indicate the results of simulations of Eq. (1a) with coupling recovery Eq. (1b), while the lines indicate the results of simulations of Eq. (1a) with the effective coupling strength Eq. (5). In both panels $\beta = 33\kappa_0$. We ran the simulations long enough for the order parameter $Z(t)$ to reach its steady state value. We calculated the time averages of $Z$ as described in Materials and methods. Error bars for the temporal standard deviations of $Z$ are smaller than the size of symbols.

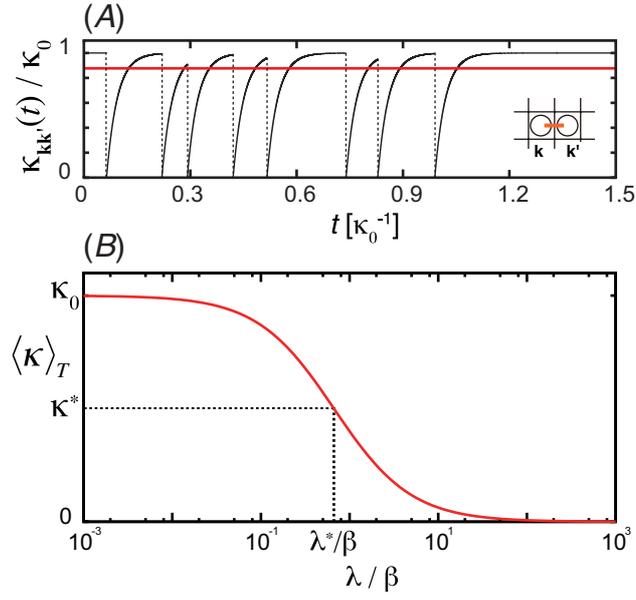

**Figure 5.** The effective coupling strength depends on the ratio of the moving rate to the coupling recovery rate $\lambda/\beta$. (A) Time series of the coupling strength between two adjacent sites, **k** and **k'**, in a two-dimensional lattice. The red line indicates the time average of the time series. (B) Dependence of the time average of the coupling strength given by Eq. (5) on the ratio $\lambda/\beta$. The horizontal broken line indicates the critical coupling strength $\kappa^*$ below which the disordered state is stable. The intersection between $\kappa^*$ and the time average of the coupling strength determines the critical moving rate $\lambda^*$.

# Supporting Information

# Optimal cellular mobility for synchronization arising from the gradual recovery of intercellular interactions

Koichiro Uriu, Saúl Ares, Andrew C. Oates and Luis G. Morelli

**Text S1**

**1. Derivation of the effective coupling strength**

Here we derive Eq. (5) in the main text, the time average of the coupling strength between two adjacent sites in a two-dimensional lattice. To illustrate how to calculate the time average of the coupling strength, we consider a time series of the coupling strength between two adjacent sites $\mathbf{k}$ and $\mathbf{k}'$ (*e.g.* $\mathbf{k} = (k, l)$ and $\mathbf{k}' = (k, l+1)$), denoted as $\kappa_{\mathbf{kk}'}(t)$ (figures S2(*A*) and (*B*)). For simplicity, we consider that these two sites are located in the bulk of a two-dimensional lattice.

Figure S2(*B*) shows an example of the time series of $\kappa_{\mathbf{kk}'}(t)$. In figure S2(*B*) the coupling strength $\kappa_{\mathbf{kk}'}(t)$ is reset to zero at times $t_0, t_1, t_2, …, t_n$, meaning that a cell in either site $\mathbf{k}$ or $\mathbf{k}'$ was replaced by another cell coming from another neighboring site at these time points (*e.g.* the cell in site $(k, l-1)$ exchanges its location with the one in site $\mathbf{k} = (k, l)$). Note that $\kappa_{\mathbf{kk}'}(t)$ is not affected by an exchange of location between the cells in sites $\mathbf{k}$ and $\mathbf{k}'$ (figure 1(*A*) in the main text). We assume that $T = t_n - t_0$ is sufficiently large so that many exchange events occur in this interval making $\kappa_{\mathbf{kk}'}(t)$ go to zero. The waiting times $\tau_i = t_i - t_{i-1}$ ($i = 1, 2, …, n$) represent the length of the interaction time during which the two cells that met at $t_{i-1}$ in site $\mathbf{k}$ and $\mathbf{k}'$ stay adjacent to each other (figure S2(*B*)). To calculate the time average of $\kappa_{\mathbf{kk}'}(t)$, we first derive the probability density function for the length of the interaction time $\tau_i$. Then, using the probability density function we calculate the time average of $\kappa_{\mathbf{kk}'}(t)$.

*S1.1 The probability density function for the length of the interaction time*

The probability that the cell in site $\mathbf{k}$ exchanges its location with one of its three neighbors other than the cell in site $\mathbf{k}'$ within the small time interval $\Delta t$ is $3\lambda\Delta t/4$, where $\lambda$ is the moving rate defined in the main text (figure S2(*A*)). The same is valid for the cell in site **k'**, so the probability that either of these two cells moves away from the other cell in the small time interval $\Delta t$ can be written as $3\lambda \Delta t/2$. Hence, the probability density function for the length of the interaction time is $f(\tau) = 3\lambda e^{-3\lambda\tau/2}/2$. The resulting average interaction time is $\langle \tau \rangle = \int_0^\infty \tau f(\tau) d\tau = 1/(3\lambda/2)$.

*S1.2 Time average and the effective coupling strength*

The time average of $\kappa_{\mathbf{kk'}}(t)$ in time interval $T = t_n - t_0$ is defined as:

$$\langle \kappa \rangle_T = \frac{1}{T} \int_{t_0}^{t_n} \kappa_{\mathbf{kk'}}(t) dt . \tag{S.1.1}$$

The integral in the right hand side of Eq. (S.1.1) can be split in $n$ terms as:

$$\langle \kappa \rangle_T = \frac{1}{T} \left\{ \int_{t_0}^{t_1} \kappa(t - t_0) dt + \int_{t_1}^{t_2} \kappa(t - t_1) dt + \ldots + \int_{t_{n-1}}^{t_n} \kappa(t - t_{n-1}) dt \right\}, \tag{S.1.2a}$$

where $\kappa(t) = \kappa_0 (1 - e^{-\beta t})$. Introducing new variables $s_i = t - t_{i-1}$ ($i = 1, 2, \ldots, n$), Eq. (S.1.2a) reads:

$$\langle \kappa \rangle_T = \frac{1}{T} \sum_{i=1}^{n} \int_0^{\tau_i} \kappa(s_i) ds_i . \tag{S.1.2b}$$

Eq. (S.1.2b) can be re-written as:

$$\langle \kappa \rangle_T = \left( \frac{n}{T} \right) \cdot \left( \frac{1}{n} \sum_{i=1}^{n} \int_0^{\tau_i} \kappa(s_i) ds_i \right). \tag{S.1.3}$$

Because we assumed that $T$ is sufficiently large, the first factor in Eq. (S.1.3) is $n/T = 1/\langle \tau \rangle = 3\lambda/2$ and the second is:

$$\frac{1}{n} \sum_{i=1}^{n} \int_0^{\tau_i} \kappa(s_i) ds_i \approx \int_0^\infty \left\{ \int_0^\tau \kappa(\tau') d\tau' \right\} \frac{3\lambda}{2} e^{-\frac{3\lambda}{2}\tau} d\tau = \int_0^\infty \left\{ \int_0^\tau \kappa_0 (1 - e^{-\beta\tau'}) d\tau' \right\} \frac{3\lambda}{2} e^{-\frac{3\lambda}{2}\tau} d\tau .$$

By substituting these equations into Eq. (S.1.3), we obtain:

$$\langle \kappa \rangle_T \approx \frac{1}{1 + 3(\lambda/\beta)/2},$$

which is Eq. (5) in the main text. Figure S2(*C*) shows the excellent agreement of Eq. (5) in the main text with the numerically calculated time averages of the coupling strength for $T = 300\kappa_0^{-1}$, which gives $n = 15 \sim 15 \times 10^4$ depending on the value of $\lambda$ used.

## 2. Effects of coupling recovery on synchronization in a model with coupling time delays

Previous studies showed that there are time delays in intercellular interactions through Delta-Notch signaling, and that these delays affect the collective dynamics of the segmentation clock [1,2]. For simplicity, in the main text we describe intercellular interactions without including time delays, and focus on the effects of coupling recovery on synchronization. Here we extend the model of the main text to include coupling delays, and study the effects of coupling recovery on synchronization in this extended model.

We introduce the time delay $\tau_d$ ($\tau_d > 0$) into Eq. (1a) in the main text as was done in previous studies [1,2]:

$$\frac{d\theta_j(t)}{dt} = \omega_d + \frac{1}{n_j}\sum_\alpha \kappa_{j\alpha}(t_{j\alpha})\sin[\theta_\alpha(t-\tau_d) - \theta_j(t)] + \sqrt{2C_d}\xi_j(t). \quad \text{(S.2.1)}$$

In Eq. (S.2.1) $\tau_d$ represents the time required for the phase information of cell $\alpha$ to reach its cell membranes and become visible to cell $j$. For simplicity, we ignore the time that cell $j$ needs to process the phase information received from cell $\alpha$. In what follows, the equation for coupling recovery Eq. (1b) in the main text, description of cell movement, and boundary conditions for phase dynamics and cell movement are the same as in the main text. When a cell moves, it carries with it its past history.

To set an initial phase history for each oscillator, its phase value at $t = -\tau_d$ is chosen randomly from a uniform distribution between 0 and $2\pi$. The phase values between $-\tau_d$ and 0 are given by:

$$\theta_j(t) = \omega_d t + \omega_d \tau_d + \theta_j(-\tau_d) \quad \text{for } -\tau_d < t \leq 0. \quad \text{(S.2.2)}$$

This initial condition represents a situation where oscillators are not coupled until the time $t = 0$ and at this time point they start to interact with each other. At $t = 0$, the coupling strength for each pair of adjacent cells is set to its maximum capacity $\kappa_0$.

We set the values of parameters within the range estimated for the zebrafish segmentation clock [1]: $\kappa_0 = 0.03$ min$^{-1}$, $\omega_d = 7.33\kappa_0$, and $\tau_d = 0.66\kappa_0^{-1}$, respectively in Eq. (S.2.1). The noise strength $C_d$ is chosen as $C_d = 0.01\kappa_0$. We numerically solved Eq. (S.2.1) with the Euler method together with the Gillespie algorithm for cell movement.

First, we consider the simple case where cells instantaneously establish intercellular interactions after movement (i.e. the coupling recovery rate $\beta$ in Eq. (1b) in the main text is infinite) to examine the effect of cellular mobility on synchronization alone under the presence of the coupling time delay (figure S3). When cells do not move ($\lambda = 0$, red triangles in figure S3(*B*)), the order parameter $Z(t)$ defined by Eq. (2) in the main text remains very close to zero even after long time. These non-mobile oscillators tend to form local checkerboard patterns where phase values between nearest neighbors are negatively correlated (figures S3(*A*) and (*C*)). These checkerboard patterns are broken at several places due to noise. Accordingly, correlations of phases are rapidly lost with increasing distance (figure S3(*C*)).

In contrast, when cells exchange their locations every $1/4\kappa_0$ on average ($\lambda = 4\kappa_0$, green circles in figure S3(*B*)), the order parameter increases with time and finally reaches a steady state value around one. These mobile oscillators achieve positive and long-range correlation of phases (figure S3(*C*)). Cellular mobility disturbs the formation of local checkerboard patterns and leads to global synchronization across the population of coupled oscillators. When the moving rate is further increased ($\lambda = 40\kappa_0$, blue squares in figure S3(*B*)), global synchronization is realized much faster. Thus, under instantaneous coupling recovery cellular mobility monotonically increases the degree of synchronization in the presence of the coupling time delay, as it does in the model without the delay.

Next, we consider the gradual coupling recovery ($\beta = 33\kappa_0$) after movement together with coupling time delays (figure S4). When $\lambda = 4\kappa_0$, mobility of oscillators still enhances the degree of synchronization compared to the case of non-mobile oscillators (green circles in figures S4(*B*) and (*C*)). Remarkably, when the moving rate becomes higher ($\lambda = 40\kappa_0$), cellular mobility hampers synchronization where neither local nor global phase order exists (figures S4(*A*) and (*C*)). Note that the value of the critical moving rate for synchronization is different from the one for the model without coupling delays, Eq. (8) in the main text. This is because the coupling time delay affects the value of the critical coupling strength together with the noise strength [3]. Nevertheless, these results indicate the existence of the optimal and critical moving rates for synchronization even in the presence of coupling time delays.

**References in Supplementary Information**

[1]     Herrgen L, Ares S, Morelli L G, Schroter C, Julicher F and Oates A C 2010 Intercellular coupling regulates the period of the segmentation clock *Curr Biol* **20** 1244-53

[2]     Morelli L G, Ares S, Herrgen L, Schroter C, Julicher F and Oates A C 2009 Delayed coupling theory of vertebrate segmentation *HFSP J* **3** 55-66

[3]     Yeung M K S and Strogatz S H 1999 Time delay in the Kuramoto model of coupled oscillators *Phys Rev Lett* **82** 648-51

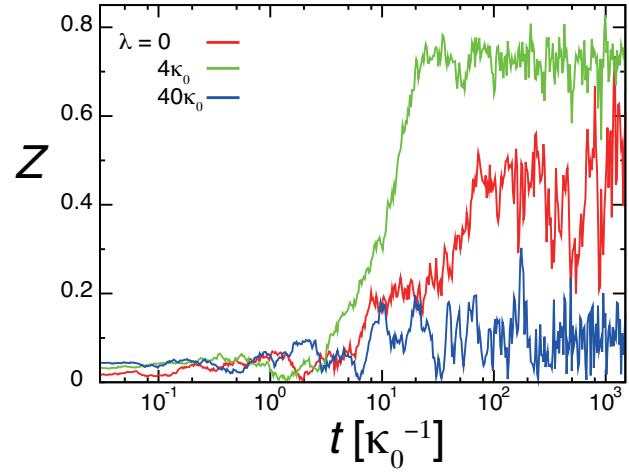

**Figure S1.** Time evolution of single trajectories of the order parameter for different moving rates $\lambda$. Parameters are $\beta = 33\kappa_0$ and $C = 0.25\kappa_0$.

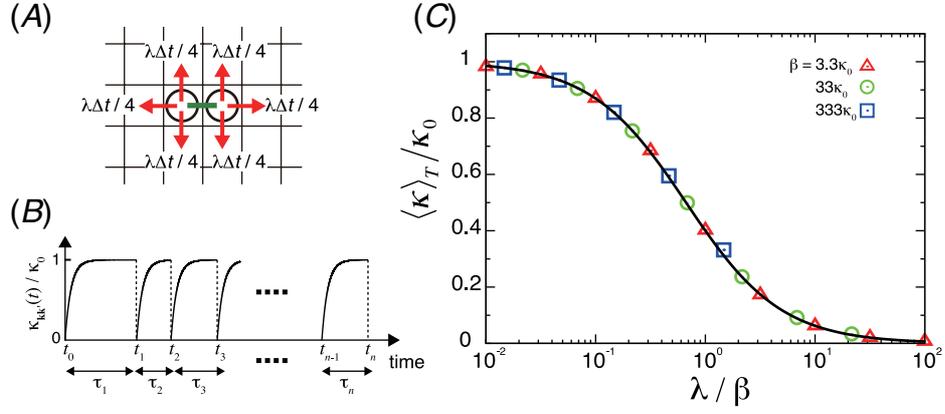

**Figure S2.** Eq. (5) in the main text agrees with the time average of the coupling strength between two adjacent sites calculated by numerical simulations. (A) The probability that one of two cells in sites $\mathbf{k}$ and $\mathbf{k'}$ moves away from the other within small time interval $\Delta t$. (B) The time series of the coupling strength between sites $\mathbf{k}$ and $\mathbf{k'}$. (C) Comparison between the time average of the coupling strength between two adjacent sites calculated by numerical simulations with different coupling recovery rate $\beta$ (symbols), and the effective coupling strength given by Eq. (5) in the main text (solid line).

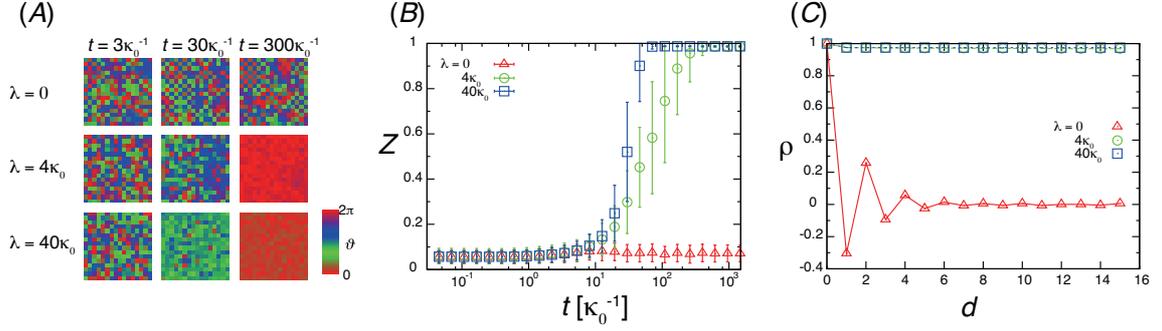

**Figure S3.** Effects of coupling delay for instantaneous coupling recovery. The degree of synchronization monotonically increases with the increase of the moving rate $\lambda$ in a model including coupling time delay, Eq. (S.2.1), when coupling recovery is instantaneous, $\beta = \infty$ in Eq. (1b). (A) Snapshots of spatial phase profiles in the two-dimensional lattice at different moving rates $\lambda$, observed in numerical simulations of Eq. (S.2.1). The phase $\vartheta$ in each site is represented by a color look-up table as indicated. (B) Time evolution of the order parameter $Z(t)$ defined by Eq. (2) in the main text for different moving rates $\lambda$. Error bars indicate standard deviations. (C) Dependence of the correlation defined by Eq. (3) in the main text on the distance $d$ between two sites. We plotted the temporal average of the correlation after its value reached a steady state. Error bars for temporal standard deviations are smaller than the size of symbols in (C). $C_d = 0.01\kappa_0$ and other parameters in Eq. (S.2.1) are as described in the supporting information.

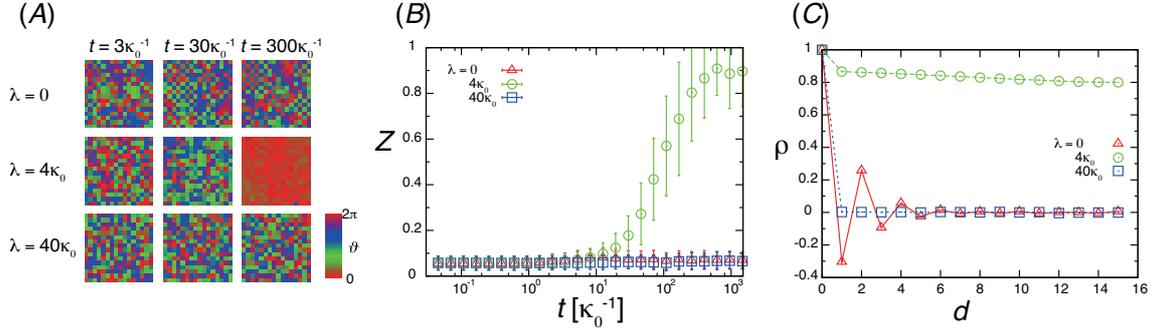

**Figure S4.** Effects of coupling delay for gradual coupling recovery. Dependence of the degree of synchronization on the moving rate $\lambda$ is non-monotonic for gradual coupling recovery ($\beta = 33\kappa_0$), in a model including coupling time delay, Eq. (S.2.1). (A) Snapshots of spatial phase profiles in the two-dimensional lattice at different moving rates $\lambda$ observed in numerical simulations of Eq. (S.2.1). The phase $\vartheta$ in each site is represented by a color look-up table as indicated. (B) Time evolution of the order parameter $Z(t)$ defined by Eq. (2) in the main text for different moving rates $\lambda$. Error bars indicate standard deviations. (C) Dependence of the correlation defined by Eq. (3) in the main text on the distance $d$ between two sites. We plotted the temporal average of the correlation after its value reached a steady state. Error bars for temporal standard deviations are smaller than the size of symbols in (C). In all panels $C_d = 0.01\kappa_0$ and other parameters in Eq. (S.2.1) are as described in the supporting information.

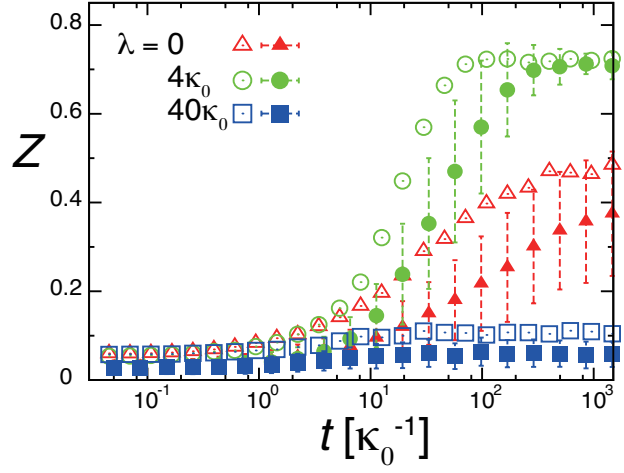

**Figure S5.** Effects of system size. Time evolution of the order parameter $Z(t)$ defined by Eq. (2) in the main text for different moving rates $\lambda$ with the system size of $32 \times 32$ (filled symbols) and $16 \times 16$ (open symbols; data shown in figure 2(B) of the main text, plotted for comparison). Error bars of filled symbols indicate standard deviations. Error bars of open symbols are shown in figure 2(B) of the main text. Parameters are $\beta = 33\kappa_0$ and $C = 0.25\kappa_0$.

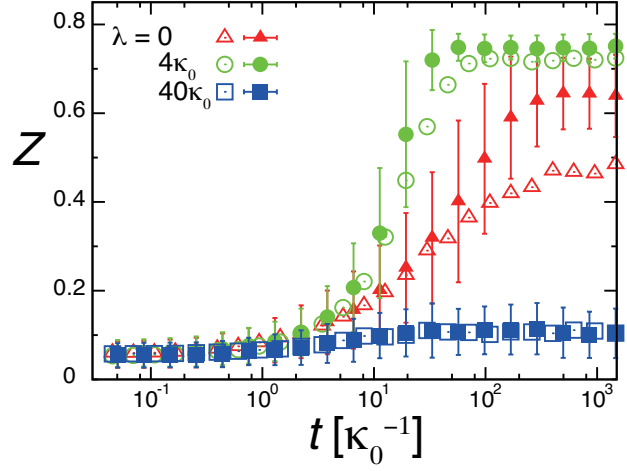

**Figure S6.** Effects of boundary conditions. Time evolution of the order parameter $Z(t)$ defined by Eq. (2) in the main text for different moving rates $\lambda$, with periodic boundary conditions (filled symbols) and open boundary conditions (open symbols; data shown in figure 2(B) in the main text, plotted for comparison). Error bars of filled symbols indicate standard deviations. Error bars of open symbols are shown in figure 2(B) of the main text. The behavior is qualitatively the same in both cases, showing that the choice of boundary condition does not affect our conclusions. For periodic boundary conditions the system size is effectively smaller, resulting in a quantitative difference for small values of the moving rate $\lambda$. Parameters are $\beta = 33\kappa_0$ and $C = 0.25\kappa_0$.

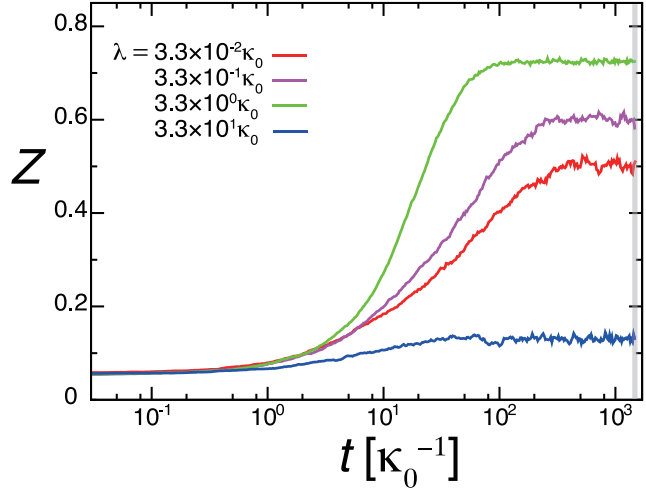

**Figure S7.** Illustration of steady state measurement of the order parameter as defined by Eq. (2) in the main text. The figure shows the time evolutions of the order parameter in simulations of Eq. (1a) with coupling recovery Eq. (1b) in the main text. The thin gray vertical band spanning from $t = 1410\kappa_0^{-1}$ to $t = 1500\kappa_0^{-1}$ indicates the time window over which we calculate the time average for this particular steady state measurement. Parameters are $\beta = 33\kappa_0$ and $C = 0.25\kappa_0$.